\def\Napoles{Departimento di Scienze Fisiche, Mostra d'Oltremare Pad. 19,
80125 Napoli, Italiy}

\def\Granadainst{Instituto de F\'{\i}sica Te\'{o}rica y Computacional Carlos I,
Facultad
de Ciencias, Universidad de Granada, Campus de Fuentenueva, Granada 18002,
Spain}

\def\Comision{Work partially supported by the DGICYT.}

\def\IAA{Instituto de Astrof\'{\i}sica de Andaluc\'{\i}a, Apartado Postal 3004,
18080 Granada, Spain}

\def\Gales{Department of Physics, University of Wales Swansea, Singleton Park, 
Swansea, SA2 8PP, U.K.}

\def\nn{\nonumber}
\def\ni{\noindent}

\def\be{\begin{equation}}
\def\ee{\end{equation}}
\def\bea{\begin{eqnarray}}
\def\eea{\end{eqnarray}}
\def\ba{\begin{array}}
\def\ea{\end{array}}

\def\z{\zeta}

\def\Gt{$\widetilde{G}\,\,$}
\def\Gtm{\widetilde{G}\,}

\def\H{{\cal H}}
\def\HT{{\cal H}_T}

\def\k{\vec{k}}

\def\L{\vec{L}}

\def\x{\vec{x}}

\def\z{\zeta}
\def\medio{\frac{1}{2}}

\def\ni{\noindent}

\def\w{\omega}

\def\hx{\tilde{x}}
\def\hy{\tilde{y}}
\def\htau{\tilde{\tau}}
\def\hPsi{\widetilde{\Psi}}
\def\hchi{\widetilde{\chi}}
\def\Gg{G_{\rm good}}
\def\varphiv{\vec{\varphi}}

\newcommand{\parcial}[1]{ \frac{\partial}{\partial #1} }

\newcommand{\XL}[1]{ {\tilde{X}}^{L}_{#1} }

\newcommand{\XR}[1]{ {\tilde{X}}^{R}_{#1} }

\documentstyle[12pt,epsf]{article}

\begin{document}



\begin{center}
{\large {\bf QUANTIZATION ON THE TORUS AND MODULAR INVARIANCE}}
\end{center}

\bigskip
\bigskip

\centerline{J. Guerrero $^{1,2,3}$, M. Calixto$^{2,4}$ and  V. Aldaya$^{1,2}$ }

\bigskip
\centerline{January 21, 1999}
\bigskip

\footnotetext[1]{\IAA}
\footnotetext[2]{\Granadainst}
\footnotetext[3]{\Napoles}
\footnotetext[4]{\Gales}

\bigskip

\begin{center}
{\bf Abstract}
\end{center}

\small

\begin{list}{}{\setlength{\leftmargin}{3pc}\setlength{\rightmargin}{3pc}}
\item The implementation of modular invariance on the torus as a phase space 
at the quantum level is discussed in a group-theoretical framework. Unlike
the classical case, at the quantum level some restrictions on the
parameters of the theory should be imposed to ensure modular invariance.
Two cases must be considered, depending on the cohomology class of the
symplectic form on the torus. If it is of integer cohomology class $n$, then
full modular invariance is achieved at the quantum level only for those
wave functions on the torus which are periodic if $n$ is even, or
antiperiodic if $n$ is odd.
If the symplectic form is of rational cohomology class $\frac{n}{r}$, a similar
result holds --the wave functions must be either periodic or antiperiodic
on a torus $r$ times larger in both direccions, depending on the parity of
$nr$. Application of these results to the Abelian Chern-Simons is discussed.

\end{list}

\normalsize

\vskip 0.25cm




\vfil\eject

\section{Introduction}

Since the pioneer work by Dirac \cite{Dirac} on the quantization of 
constrained systems, a lot of work has been done on this subject, and 
plenty of methods have been developed to face this interesting and, many 
times, difficult problem. Roughly speaking, the different methods can be 
classified into two types, depending on whether the quantization of the 
corresponding unconstrained system is first performed and then the constraints
imposed at the quantum level (``quantize-fist'' method) or  the constraints 
are first imposed and then the quantization of the resulting ``reduced'' 
system is performed (``constrain-first'' method). An example of the former is 
given by the abovementioned paper by Dirac \cite{Dirac}, while  the latter 
was originated by the work of Faddeev \cite{Faddeev}. Many other procedures 
derive from these two, adapted to the properties of the particular system 
under consideration. Thus, for instance, the BRST quantization is a 
``quantize-first'' technique adapted to the covariant quantization of gauge 
invariant systems \cite{BRST}. Also, the method proposed by Ashtekar 
\cite{Ashtekar} was designed to simplify the form of the quantum constraints 
in quantizing Gravity. Alternatively, symplectic or 
Marsden-Weinstein reduction \cite{Marsden-Weinstein} is a specific technique 
developed to obtain a reduced classical phase space, which must be the 
starting point for (some sort of) Geometric Quantization \cite{Woodhouse}.    

The main drawback of the ``constrain-first'' method lies in that the classical
phase space could be not properly defined as a differential manifold or, even
more, the classical equation of motion might have no general solution. In 
addition, all the problems that Geometric Quantization encounters in 
dealing with non-trivial phase spaces must be considered (anomalies, i.e the
lack of invariant polarizations, the search for operators compatible with
the polarization, etc.).

The troubles with the ``quantize-first'' methods appears in the implementation
of the quantum constraints; only quadratic constraints can be directly 
imposed due to normal-order ambiguities. Besides, finding the operators
which preserve the quantum constraints is a non-trivial problem.

In \cite{Comm1} a method for studying quantum systems with constraints on a 
group-theoretical framework, Algebraic Quantization on a Group
(AQG),  was introduced. AQG is a ``quantize first'' method in which both the 
unconstrained systems and the constraints are supposed to be dealt in a 
group setting. This could seem at first instance a severe restriction but, 
in practice, most of the interesting cases can be treated with this formalism, 
and the advantages it provides are numerous. In particular, there are no 
ambiguities in the imposition of quantum constraints 
(even for non-polynomical ones), and there is a 
characterization for the operators that preserve the quantum constraints.

Another advantage of AQG is the possibility of implementing the non-trivial
topology of the configuration or phase space as contraints, taking into account
that these are discrete transformation, which can be easily addressed in this 
formalism.

In \cite{frachall} the quantization of the Heisenberg-Weyl (H-W) group with
constraints was considered and the particular case of the H-W group on the
torus was studied. Now, we wish to implement modular invariance
on the torus at the quantum level. In general, the modular invariance of a 
conformal theory formulated on a Riemannian surface of genus $g$, $\Sigma_g$, 
refers to the quotient group $Diff(\Sigma_g)/Diff_0(\Sigma_g)$, where 
$Diff(\Sigma_g)$ is the group of diffeomorphisms of $\Sigma_g$ and the 
subscript $0$ designs the normal subgroup of diffeomorphisms connected to the 
identity (see, for instance, \cite{Martin,Nakahara}). 

Clearly, modular transformations on the
torus are the $SL(2,Z)$ subgroup of the group $SL(2,R)\approx Sp(2,R)$ of 
linear symplectic transformations of the plane that preserves the torus.
Therefore we can implement them in the formalism of Algebraic
Quantization on a Group by considering the Schr\"{o}dinger group
(or Weyl-Symplectic group, see \cite{chorri})
$WSp(2,R)$ as the symmetry group of the unconstrained system and imposing the
appropriate constraints to obtain a
torus as the (reduced) symplectic manifold, pretty much in the same manner
as in \cite{frachall}. Then we expect to obtain modular
transformations as good operators, i.e. those preserving the Hilbert
space of wave functions satisfying the constraints. However, to
obtain full modular invariance, we must impose some restrictions on the 
parameters of the theory. As in \cite{frachall}, three different cases
should be considered, depending on the cohomology class of the symplectic
form on the torus, which can be either integer, fractional or irrational.
Only the integer and fractional cases will be considered here, since the
irrational one requires techniques from Non-Commutative Geometry \cite{Connes}
and lies beyond the scope of this paper.

These results are applied to $2+1$D Abelian Chern-Simons theory and compared
with the ones obtained in the literature.


The present paper is organized as follows: In section \ref{Schrodinger} we 
study the
Schr\"{o}dinger group without constraints and compute the metaplectic (or
spinor) representation with
the help of a higher-order polarization. Section \ref{Schrodinger-toro} is 
devoted to the
determination of the constrained Hilbert space and good operators when the
phase-space is constrained to be a torus. Two cases are
considered, the one for which the symplectic form on the torus is of
integer cohomology class $n$ (section \ref{integer}), where full 
modular invariance
is obtained only when the wave functions are periodic for $n$ even or
antiperiodic for $n$ odd, and the case of symplectic form
of rational cohomology class $\frac{n}{r}$ (section \ref{fractional}),
where full modular invariance is obtained only when the
wave functions are periodic for $nr$ even or antiperiodic for $nr$  odd.
Here periodicity and antiperiodicity are understood in a torus which is r times
larger in both directions. Finally, section \ref{C-S} is devoted to the 
application of 
our study to $2+1$D Abelian Chern-Simons theory.

In a separate Appendix, we study the representations of the
subgroup $T$ both for the integral and fractional case.

\section{Algebraic Quantization on a Group}
\label{AQG}

Algebraic Quantization on a Group (AQG) (see \cite{Comm1,frachall}) is a 
group-theoretical procedure developed for quantizing systems with constraints 
(both first- and second-class) in a first-quantize-then-constrain basis. The 
starting point
is the group \Gt of quantum symmetries of the unconstrained system, which
is a central extension by $U(1)$ of the group $G$ of classical
symmetries of the unconstrained system. From \Gt, a subgroup $T$, called
the structure group,
is selected for defining the constraints. For convenience, $T$ is chosen to
include the $U(1)$ subgroup of the central extension, which accounts for
the phase-invariance of Quantum Mechanics ($U(1)$-equivariance), in
such a way that $\Gtm/T$ is the classical reduced phase-space of the
constrained system\footnote{To be precise, \Gt contains in general
symmetries without symplectic content, like time translations or rotations,
so that $\Gtm/T$ is the reduced pre-symplectic manifold of the constrained
system.}.

 The quantum Hilbert space $\HT$ for the constrained system is defined
by selecting, from the Hilbert space $\H$ associated with a unitary
irreducible representation $U(\Gtm)$ of \Gt, those wave functions that
transform irreducibly under a given unitary irreducible representation
$D(T)$ of $T$.
We shall say that these wave functions satisfy the $T$-function condition
(or $T$-equivariance condition), which has the general form:
\be
\Psi^{\alpha}(g_T*g) = D^{\alpha}(g_T) \Psi^{\alpha}(g)\,,\,\,\,
\forall g_T\in T\,, \label{T-funcion}
\ee

\ni where the index $\alpha$ in $D$ ranges over the set $\hat{T}$, the
Pontryagin dual of $T$ --that is, the set of all unitary irreducible
representations of $T$. Precisely stated, $\alpha$ will be allowed to vary
along the subset $\hat{T}_U\subset \hat{T}$ of
those representations which are contained in the restriction of $U(\Gtm)$ to
$T$; otherwise the constraints would be inconsistent and the constrained Hilbert
space $\HT$ would be trivial. In particular, the representation
$D^{\alpha}$, when restricted to the
subgroup $U(1)\subset T$, should be the natural (faithful) representation
of $U(1)$, $D^{\alpha}(\z) = \z\,,\,\forall \z\in U(1)$. That is, the
$T$-equivariance condition must contain the $U(1)$-equivariance condition.
Complex functions on the group satisfaying the $T$-equivarance condition
can be identified with sections of the vector bundle associated with the
principal bundle $T\rightarrow\Gtm\rightarrow \Gtm/T$ through the 
representation $D^{\alpha}$ of $T$ \cite{Landsman}.

Both the unitary irreducible representations $U(\Gtm)$ and $D(T)$ can be
obtained, for instance, by using the Group Approach to Quantization (GAQ)
technique (see \cite{Comm1} and references therein), which uses the method
of polarizations (see below) to reduce the left-regular
representation of the group acting on $U(1)$-equivariant complex functions
on the group \Gt.

 An important concept that we are forced to introduce is the notion of {\it good
operators}, defined as those preserving the constrained
Hilbert space $\HT$. It is clear that, since $\HT$ is in general smaller than
$\H$, not all operators in \Gt will preserve it; otherwise the representation
$U(\Gtm)$ would be reducible. It is difficult to give a general
characterization of these
operators (for instance, there can be operators preserving $\HT$ which
belong neither to \Gt nor to its enveloping algebra,
escaping to any algebraic or differential characterization), but we can find
all good operators in \Gt simply by considering the {\it little group} of the
representation $D^{\alpha}(T)$ of $T$ --that is, the subgroup $\Gg$ of elements
$g_g$ that send the representation $D^{\alpha}(T)$ to an equivalent one
under the adjoint action:
\be D_{g_g}^{\alpha}(g_T)\equiv D^{\alpha}(g_g*g_T*g_g^{-1}) \approx
D^{\alpha}(g_T)\,\,\,\,\, \forall g_T\in T\,,\,\,\forall g_g\in \Gg\,. \label{pequegno}
\ee

\ni Note that this definition generalizes the (sufficient) ones given in
\cite{Comm1} and \cite{frachall}. For instance, in the case
in which the representation $D^{\alpha}(T)$ is one-dimensional (in particular,
if $T$ is Abelian), the definition above gives
$D_{g_g}^{\alpha}(T)=D^{\alpha}(T)$, and the sufficient condition given
in \cite{frachall},
\be
[\Gg, T] \subset {\rm ker} D^{\alpha}(T) \label{guenos}\,,
\ee

\ni also proves to be necessary. This characterization reproduces the standard
one for the case of first-class
constraints, for which $T=C\times U(1)$, where $C$ is the
subgroup of constraints ($U(1)$ only accounts for the phase-invariance of
Quantum Mechanics). If we choose for $C$ the trivial representation (for
$U(1)$ the natural representation must always be chosen), then
\be
[\Gg,C]\subset C \,.
\ee

\ni This condition gives $\Gg$ as the normalizer of the constraints, as it is
usually the case (see, for instance, \cite{Ashtekar}). However, if a non-trivial
representation of $C$ is
chosen, the subgroup of good operators can be smaller
that the normalizer
of the constraints, revealing a strong dependence of $\Gg$ on the representation
$D^{\alpha}(T)$ of $T$ and, therefore, we should use the more precise notation
$\Gg^{\alpha}$ for the subgroup of operators preserving the reduced Hilbert
space $\HT^{\alpha}$. Note that, from the very definition of little group,
$\Gg^{\alpha}\subset N_T,\,\,\forall \alpha\in \hat{T}_U$, where $N_T$ is the
normalizer of T in \Gt, so that the appropriate place to look for good operators
will be in $N_T$.

It is useful to examine the case in which $C$ is an invariant subgroup of \Gt and we
choose $D(T)$ to be the restriction of $U(\Gtm)$ to $T$ (or $U(\Gtm)$ to be
the induced representation by $D(T)$). Then the constraints are trivial; i.e.,
they do not imply additional restrictions on the wave functions, and the
constrained and unconstrained Hilbert spaces coincide. Moreover, the subgroup
of good operators turns out to be the whole \Gt. In this case, $C$ is called a
gauge group (see \cite{gauge}).

A separate study is warranted by the case when $T$ cannot be written as
$C\times U(1)$, for instance when $T$ is a non-trivial central extension of
$C$ by $U(1)$. In this case, $C$ contains canonically conjugated variables,
and the constraints are of {\it second class}. This case, also contemplated
in \cite{Comm1,frachall}, will be studied in section \ref{fractional}.

It should be noted that the same programme can be carried out considering the
Lie algebras $\tilde{\cal G}$ of \Gt and ${\cal T}$ of $T$, when these are
simply connected groups. In this case, the treatment becomes simpler, since the
representations $dU(\tilde{\cal G})$ and $dD({\cal T})$ are easier to obtain.
In general, however, the treatment is more involved, not only because
the good operators can lie in the enveloping algebra, but also because
the constraints themselves can be defined through higher-order differential
equations \cite{conforme}. But all these cases can be handled with a direct
generalization of AQG.

Thus, AQG can be applied to constrained systems, irrespective of the type
(first- or second-class) of constraints. Some examples of application of AQG
can be found in \cite{Comm1}, where parity in a two-particle system was
introduced to obtain both bosonic and fermionic quantizations, and
diffeomorphisms constraints to obtain the bosonic string.

Other interesting examples for applying AQG are those systems in which the
configuration or phase-spaces are multiply connected and the group \Gt of
quantum symmetries of the simply connected counterpart (universal covering)
is known.
If $P$ is a multiply connected phase space which is homogeneous under a group
$G$ of symmetries, then $P$ is locally diffeomorphic to a coadjoint orbit of
$G$, or to a coadjoint orbit of a central extension of $G$ by $U(1)$ or $R$,
\Gt \cite{Kirillov}.
For the first case, if $H$ is the isotropy group of $P$, $G/H$ is locally
diffeomorphic to $P$. If we choose $G$ appropriately (taking coverings, if
necessary) in such a way that $G/H$ is simply connected, then $P$ is the
quotient of $G/H$ by $\pi_1(P)$, the first homotopy group of $P$.
For the cases in which $P$ is locally diffeomorphic to a coadjoint orbit of
a central extension \Gt of G, and if \Gt is choosen (taking coverings) in
such a way that this orbit is simply connected, then $P$ is the quotient of
$\Gtm/H$ by $\pi_1(P)\times U(1)$ (or $R$). Then $C=\pi_1(P)$ and
$T=C\times U(1)$.

However, if $P$ is not the cotangent bundle of any configuration space
(as, for instance, the sphere or the torus as symplectic
manifolds), then it could well happen that $\pi_1(P)$, as a subset of
\Gt (we should not forget that all operations of taking quotients are
done in \Gt, and therefore we must consider the embeding of $\pi_1(P)$ in
\Gt, and this could be not a group), contains canonically conjugated pairs.
In this case, $T$ is a central extension of $C$ by $U(1)$ and the
constraints are of second class. However, if the representation $D$ of $T$
is finite-dimensional (see Appendix A), even though $T$ defines
second-class constraints, the treatment follows as though they were
first-class, yet non-Abelian.


\section{The Schr\"{o}dinger group}
\label{Schrodinger}

As mentioned in the Introduction, we shall replace the H-W group used 
in \cite{frachall} with the Schr\"odinger group, which coincides with the 
Weyl-Symplectic group $WSp(2,R)$ in 1 dimension. Is was first studied by
Niederer \cite{Niederer} as the maximal kinematical invariance group of the
Schr\"odinger equation with general quadratic potential. The complete
classification of its unitary irreducible representations was given in
\cite{Perroud}. Mathematically it can be obtained from the Galilei (or from 
the Newton) group by
replacing the time parameter with the three-parameter group $SL(2,R)$. The
interest of the $SL(2,R)$ group in the present work lies in that it 
constitutes the maximal finite subgroup of the diffeomorphisms (in fact 
symplectomorphisms) group of the phase space $R^2$ (see \cite{symplin}, where 
some physical meaning is given to the representations considered 
``unphysical'' in \cite{Perroud}).

To perform a global-coordinate treatment of the problem, we shall
start by considering matrices $S\in GL(2,R)$ instead of $SL(2,R)$, and the
condition for these matrices to belong to $SL(2,R)$ will appear naturally.
A group law for the Schr\"{o}dinger group can be written as:
\bea
\x{}\ ''&=& \x\ ' + \frac{S'}{|S'|^{1/2}}\x \nn \\
S'' &=& S' S \\
\z''&=&\z'\z \exp\frac{im\w}{2\hbar}\left[
 \frac{-A'x_2'x_1-B'x_2'x_2+C'x_1'x_1+D'x_1'x_2}{|S'|^{1/2}}\right]\,,\nn
\eea

\ni where $\x=(x_1,x_2) \in R^2$,
$S=\left(\ba{lr} A&B\\C&D\ea\right)\in GL(2,R)$,  $|S|\equiv AD-BC$ and
$\frac{m\w}{\hbar}$ is an adimensional constant parametrizing the
central extensions of the H-W group (we write it in this form for later
convenience). The factor $|S'|^{-1/2}$ in the semidirect action of
$GL(2,R)$ is needed in order to have a proper central extension.

Let us quantize this system using GAQ, whose principal
ingredients will be introduced as needed (see \cite{Comm1} for details).
From the group law, the left-invariant vector fields associated with the
coordinates $x_1,x_2,A,B,C,D,\z$,
\bea
\XL{x_1} &=& |S|^{-1/2}\left[ A\parcial{x_1} + C\parcial{x_2}
              + \frac{m\w}{2\hbar}(-Ax_2+Cx_1)\Xi\right] \nn \\
\XL{x_2} &=& |S|^{-1/2}\left[ D\parcial{x_2} + B\parcial{x_1}
              + \frac{m\w}{2\hbar}(-Bx_2+Dx_1)\Xi\right] \nn \\
\XL{A} &=& A\parcial{A} + C\parcial{C}  \nn \\
\XL{B} &=& A\parcial{B} + C\parcial{D}  \\
\XL{C} &=& B\parcial{A} + D\parcial{C} \nn \\
\XL{D} &=& B\parcial{B} + D\parcial{D} \nn\\
\XL{\z}&=& i\z\parcial{\z}\equiv \Xi\,, \nn
\eea

\ni as well as the right-invariants ones,
\bea
\XR{x_1} &=& \parcial{x_1} + \frac{m\w}{2\hbar}x_2\Xi \nn\\
\XR{x_2} &=& \parcial{x_2} - \frac{m\w}{2\hbar}x_1\Xi \nn\\
\XR{A} &=& A\parcial{A} + B\parcial{B}+\medio x_1\parcial{x_1}
          -\medio x_2\parcial{x_2} \nn \\
\XR{B} &=& C\parcial{A} + D\parcial{B} + x_2\parcial{x_1} \\
\XR{C} &=& A\parcial{C} + B\parcial{D} + x_1\parcial{x_2} \nn \\
\XR{D} &=& C\parcial{C} + D\parcial{D}-\medio x_1\parcial{x_1}
           + \medio x_2\parcial{x_2} \nn \\
\XR{\z} &=& \Xi\,,
\eea

\ni can be obtained. The commutation relations for the (left) Lie algebra are:
\be
\ba{lcl}
\left[ \XL{A},\XL{B} \right] &=& \XL{B} \\
\left[ \XL{A},\XL{C} \right] &=& -\XL{C} \\
\left[ \XL{A},\XL{D} \right] &=& 0 \\
\left[ \XL{B},\XL{C} \right] &=& \XL{A}-\XL{D} \\
\left[ \XL{B},\XL{D} \right] &=& \XL{B} \\
\left[ \XL{C},\XL{D} \right] &=& -\XL{C} \\
\left[ \XL{x_1},\XL{x_2} \right] &=& \frac{m\w}{\hbar}\Xi \\
\left[ \XL{A},\XL{x_1} \right] &=& \medio\XL{x_1}
\ea
\ba{lcl}
\left[ \XL{A},\XL{x_2} \right] &=& -\medio\XL{x_2} \\
\left[ \XL{B},\XL{x_1} \right] &=& 0 \\
\left[ \XL{B},\XL{x_2} \right] &=& \XL{x_1} \\
\left[ \XL{C},\XL{x_1} \right] &=& \XL{x_1} \\
\left[ \XL{C},\XL{x_2} \right] &=& 0\\
\left[ \XL{D},\XL{x_1} \right] &=& -\medio\XL{x_1} \\
\left[ \XL{D},\XL{x_2} \right] &=& \medio\XL{x_2}\,.
\ea
\ee

From these commutation relations we see that two linear combinations of
vector fields can be introduced, $\XL{A}-\XL{D}$ and $\XL{A}+\XL{D}$ (the same
for the right-invariant vector fields), in such a way that $\XL{A}+\XL{D}$
is a central generator, which is also horizontal (see below), and therefore
is a gauge generator (see \cite{gauge}). In fact, it coincides with its right
version, as is always the case for a central generator.

We define the Quantization 1-form $\Theta$ as the vertical component
(dual to the vertical generator $\Xi$, in this basis) of the canonical
1-form of the Lie algebra:
\be
\Theta =\frac{m\w}{2\hbar}(x_2dx_1-x_1dx_2) + \frac{d\z}{i\z} \,.
\ee

\ni The 2-form $d\Theta$ defines a presymplectic form on \Gt, and
its value at the identity, $\Sigma=d\Theta|_e$, is a 2-co-cycle of
the Lie-algebra, and it can be used to characterize the central
extension (when the group \Gt is simply connected). A subalgebra is said to
be {\it horizontal} if it lies in the
kernel of $\Theta$. The {\it characteristic subalgebra} is defined as
${\cal G}_{\Theta}={\rm Ker}\Theta\cap{\rm Ker}d\Theta$, and in this case
it has the form:
\be
{\cal G}_{\Theta} = <\XL{A}+\XL{D}, \XL{A}-\XL{D}, \XL{B},\XL{C} >\,.
\ee

\ni Note that $d\Theta/({\rm Ker} d\Theta)$ defines a true symplectic form in
$R^2$.

We define the representation $U(\Gtm)$ of \Gt to be given by the left regular
representation on complex
wave functions over \Gt, satisfying the $U(1)$-function condition
$\Xi\Psi=i\Psi$ (phase invariance of Quantum Mechanics). This representation
is obvioulsy reducible, and additional restrictions should be imposed on the
wave functions in order to obtain an irreducible representation. These are
accomplished by the polarization ${\cal P}$,
defined as a maximal horizontal left subalgebra of $\tilde{\cal G}$. The
condition $X^L\Psi=0, \forall X^L\in {\cal P}$ leads, in most of the cases, to an
irreducible representation $U(\Gtm)$ acting on the Hilbert space
$\H$ of complex
polarized functions on the group satisfying the $U(1)$-function condition.


 However, there are groups, called anomalous (see \cite{chorri}), for
which this representation $U(\Gtm)$ so obtained is not irreducible, and
a generalization of the concept of polarization is required for them. This
task is acomplished by means of higher-order polarizations (see
\cite{chorri,virazorro,marmo}), which admit elements of the left enveloping
algebra to enter into them.

The system we are studying is an example of anomalous system
(see \cite{chorri,marmo}), and a higher-order
polarization is required to obtain an irreducible representation. There are
essentially two of them\footnote{There are another two, if we allow for complex
coordinates, but all of them lead to equivalent representations}, given by:
\bea
{\cal P}^{HO}&=&<\XL{A}+\XL{D},\XL{A}-\XL{D}-
   \frac{i\hbar}{2m\w}\left(\XL{x_1}\XL{x_2}+\XL{x_2}\XL{x_1}\right),
\XL{B}+\frac{i\hbar}{2m\w}\left(\XL{x_1}\right)^2, \nn \\
& & \XL{C}-\frac{i\hbar}{2m\w}\left(\XL{x_2}\right)^2,
\XL{x_1} \hbox{  or  } \XL{x_2}>\,.
\eea

If we choose, for instance, $\XL{x_1}$ to be in the polarization, the
polarization equations are:
\bea
\left(\XL{A}+\XL{D}\right)\Psi&=&0 \nn\\
\XL{B}\Psi&=&0 \nn\\
\left(\XL{A}-\XL{D}\right)\Psi&=&-\medio\Psi \\
\XL{x_1}\Psi&=& 0 \nn \\
\XL{C}\Psi &=& \frac{i\hbar}{2m\w}\left(\XL{x_2}\right)^2\Psi\,, \nn
\eea

The first of these equations has as solutions those complex wave functions on
the group $GL(2,R)$ which are defined on $SL(2,R)$, as expected. Therefore,
the solutions of this equation have the form:
\be
\Psi=\Psi(a,b,c,d,x_1,x_2)\,,
\ee

\ni where $a\equiv \frac{A}{\sqrt{AD-BC}},b\equiv \frac{B}{\sqrt{AD-BC}},
c\equiv \frac{C}{\sqrt{AD-BC}}$ and $d\equiv \frac{D}{\sqrt{AD-BC}}$, with
$ad-bc=1$.

To proceed further in solving the polarization equations, it is convenient to
introduce local charts on $SL(2,R)$. We choose them to be the ones defined
by $a\neq 0$ and $c\neq 0$, respectively\footnote{Certainly they really
correspond to four contractible charts: $a>0,a<0$ and $c<0,c>0$, but the
transition functions between each pair of these charts are trivial, so we
shall consider them as only one chart.}. The first chart contains the identity
element $I_2$ of $SL(2,R)$, and the second contains
$J\equiv\left( \ba{cc}0&1\\-1&0\ea\right)$.

The solutions to the the polarization equations are given by:
\begin{itemize}
\item For $a\neq 0$:
\be
\Psi=\z a^{-1/2} e^{\frac{im\w}{2\hbar}xy}\chi(\tau,y)\,, \label{polarizadas1}
\ee

\ni where $x\equiv x_1,\ y\equiv x_2-\tau x_1$ and $\tau\equiv \frac{c}{a}$,
with $\chi$ satisfying the Schr\"{o}dinger-like equation
\be
\frac{\partial\chi}{\partial\tau} =
\frac{i\hbar}{2m\w}\frac{\partial^2\chi}{\partial y^2}\,.
\ee

\item For $c\neq 0$:
\be
\hPsi=\z c^{-1/2} e^{-\frac{im\w}{2\hbar}\hx\hy}\hchi(\htau,\hy)\,,
\ee

\ni where $\hx\equiv x_2,\ \hy\equiv x_1-\htau x_2$ and
$\htau\equiv \frac{a}{c}$,
with $\hchi$ satisfying the Schr\"{o}dinger-like equation
\be
\frac{\partial\hchi}{\partial\htau} =
-\frac{i\hbar}{2m\w}\frac{\partial^2\hchi}{\partial \hy^2}\,.
\ee
\end{itemize}

The element $J$ represents a rotation of $\frac{\pi}{2}$ in the plane
$(x_1,x_2)$, and takes the wavefunction from one local chart to the
other\footnote{In fact, up to a factor, $J$ represents the Fourier transform
passing from the $x_1$ representation to the $x_2$ representation.}.
Obviously, $J^4=I_2$, but acting with $J$ on
the wavefunctions we obtain:
\be
\Psi(J*g)=(-1)^{1/4}\hPsi(g)\,,
\ee

\ni from which the result $\Psi(J^4*g)=-\Psi(g)$ follows, that is, the
representation obtained for the subgroup $SL(2,R)$ is two-valued. This
representation is the well-known {\it metaplectic} or {\it spinor
representation}. The
metaplectic representation is for $SL(2,R)$ as the $\medio$-spin representation
is for $SO(3)$ (see \cite{Folland} and references therein, and also
\cite{Kirillov}). We refer the reader to \cite{symplin} for a detalied
study of the Schr\"{o}dinger group, including the non-anomalous
representations and a physical interpretation for them.

\section{The Schr\"{o}dinger group on the torus}
\label{Schrodinger-toro}

Once we have obtained the polarized wave functions and therefore fixed the
unitary and irreducible representation $U(\Gtm)$ of \Gt and the unconstrained
Hilbert space $\H$, we have to impose the appropriate constraints to reduce
the phase space to a torus. This task is achieved by the structure group
$T$, which is a fibre bundle with base
$\Gamma_{\L}\equiv \{e_{\k}, \k\in Z\times Z\}$ and fibre $U(1)$, where
$e_{\k}$ are
translations of $\x$ by an amount of $\L_{\k}\equiv (k_1L_1,k_2L_2)$, in
such a way that \Gt$/T$ is essentially the torus.\footnote{As was
commented before,
$\Gtm/T$ in this case is a presymplectic manifold, which, once the kernel
of the (pre)symplectic form $d\Theta$ (containing the $SL(2,R)$ subalgebra)
is removed, turns out to be a torus.} The fibration of $T$ by $U(1)$ depends on
the values of $m, \w, L_1$ and $L_2$, and is, in general, non-trivial.

The following task is to obtain the irreducible representations of
$T$. These are studied in detail in Appendix A, and here we shall report only
the main results. The form of the representations of $T$ depends strongly
on its structure as $U(1)$ bundle with base $\Gamma_{\L}$ (which plays
the role of constraints $C$), and this is
determined by the character of the adimensional parameter
$\frac{m\w L_1L_2}{2\pi\hbar}$, in such a way that:
\begin{itemize}
\item[$i)$] {\sc integer case}: $\,\,\frac{m\w L_1L_2}{2\pi\hbar} = n \in Z$.
 In this case $T$ is Abelian, $T=\Gamma_{\L} \times U(1)$, and therefore
all its representations are 1-dimensional.
\item[$ii)$] {\sc fractional case}: $\,\,\frac{m\w L_1L_2}{2\pi\hbar}=
\frac{n}{r}\,,$ where $n$ and $r$ are relative prime integers
(with $r > 1$). In this case $T$ is not Abelian, but its
representations are of finite dimension.
\item[$iii)$] {\sc irrational case}: $\,\,\frac{m\w L_1L_2}{2\pi\hbar}= \rho$,
where $\rho$ is an irrational number. In this case, $T$ is not Abelian and
possesses representations (the ones which are compatible with the
$U(1)$-function condition) of infinite dimension. 

\end{itemize}

The irrational case will not be considered here, since its study requires 
techniques from Non-Commutative Geometry \cite{Connes}, and therefore lies
beyond the scope of the present work. The most interesting properties of
this case, in particular, of the group $C^*$-algebra generated by the 
elements of $T$, denoted {\it irrational rotation algebra}, is that it is not 
a type I algebra \cite{Rieffel} (in fact it is a type II$_{\infty}$ algebra).

Although normally the integer and fractional cases are used in physical 
applications, like Abelian Chern-Simons theory (see section \ref{C-S}) or the
Quantum Hall Effect (Integer and Fractional), there exists works 
\cite{Bellissard} where the irrational case has 
been use to study the Quantum Hall Effect, using techniques of $C^*$-algebras  
and cyclic cohomology to explain the integrality of the conductance on the
Quantum Hall Effect (see also \cite{Connes}).


\subsection{The Integer case}
\label{integer}

We shall consider first the integer case,
for which $\frac{m\w L_1L_2}{2\pi \hbar} = n \in Z$ and the structure group
is $T=\Gamma_{\L} \times U(1)$, $\Gamma_{\L}$ being a subgroup isomorphic to
$Z\times Z$. This case leads to a symplectic form on the torus of integer
cohomology class $n$ (and therefore the torus is quantizable according to
Geometric Quantization),
and $n$ can be interpreted as the Chern number of a $U(1)$-bundle
over the torus (see \cite{frachall}).

The representations of $T$ (compatible with the $U(1)$-function condition)
for the integer case are easily computed (see
Appendix A), and have the form:
\be
D^{\varphiv}(I_2,k_1L_1,k_2L_2,\z)=\z e^{i(\varphi_1k_1+\varphi_2k_2)}
e^{-i\pi nk_1k_2}\,,
\ee

\ni where $\varphi_1,\varphi_2\in [0,2\pi)$ parameterize the inequivalent
representations of the subgroup $\Gamma_{\L}\approx Z\times Z$. They are the
analogue of vacuum angles in Quantum Chromodynamics (see, for instance,
\cite{Poly}).

 The $T$-function conditions are written as
$\Psi^{\varphiv}(g_T*g)=
 D^{\varphiv}(g_T)\Psi^{\varphiv}(g),\ \forall g_T\in T$. They can be interpreted
as periodic boundary conditions, selecting those
wave functions in $\H$ which are quasi-periodic; i.e., picking up a phase
$e^{i\varphi_1}$ when translated by $L_1$ and $e^{i\varphi_2}$ when translated
by $L_2$. This condition reduces to
$\Psi^0(g_T*g)=\z\Psi^0(g)$ if the trivial representation for $\Gamma_{\L}$ is
chosen (strictly periodic boundary conditions). As in \cite{frachall}, the
rest of non-equivalent representations
can be obtained by acting with those finite translations which are not good
operators. We are not interested in their explicit form, so we refer the
interested reader to \cite{frachall} for the details of the computations.

The solutions to the $T$-function condition for the trivial representation
are those functions $\Psi$ of the
form (\ref{polarizadas1}) for which $\chi(\tau,y)$ is of the form:
\be
\chi^0(\tau,y)=\sum_{k=0}^{n-1} a_k \Delta_k^0(\tau,y)\,,
\ee

\ni with
\be
\Delta_k^0(\tau,y)=e^{i2\pi k(\frac{y}{L_2}-\frac{\delta}{2n}\tau k)}
\sum_{q\in Z} e^{i2\pi n q[\frac{y}{L_2} - \frac{\delta}{2n}\tau(k^2+nq)]}\,,
\ee

\ni $\delta\equiv \frac{L_1}{L_2}$ and $a_k,\,k=0,1,\ldots,n-1$ being
arbitrary coefficients. We can write these in a
form which resembles the one obtained in \cite{frachall}, where we considered
only the Heisenberg-Weyl subgroup:
\be
\Delta_k^0(\tau,y)=e^{i\frac{\delta L_2^2\tau}{4\pi n}\frac{d^2}{dy^2}}
\Delta_k^0(y)\,,
\ee

\ni with $\Delta_k^0(y)=e^{i2\pi k \frac{y}{L_2}}
\sum_{q\in Z}\delta(y- \frac{q}{n}L_2)$.

For the local chart at $J$ ($c\neq 0$), we could follow the same procedure
or simply transform the wave function acting with $J$. The result obtained
is completely analogous to the one obtained in the local chart at the identity.
Therefore, the constrained Hilbert space $\HT$ is finite dimensional, with
a basis of $n$ independent functions, $\{\Delta^0_k\}_{k=0}^{n-1}$.

Now we have to compute the {\it good operators}, those preserving the Hilbert
space $\HT$ of polarized wave functions verifying the $T$-function condition.
We should look for good operators in the normalizer of $T$ in \Gt. In this case
(this result is also valid for the fractional case), we have:
\be
N_T = \left\{ (\left(\ba{lr} a&b\\ c&d\ea\right),x_1,x_2,\z) \in \Gtm\,/ \,
a,\,b\delta^{-1},\,c\delta,\,d \in Z,\,\, x_1,\,x_2\in R,\,\,\z\in U(1)\right\}\,,
\ee

\ni which implies that $N_T$ is the semidirect product of $SL(2,Z)$ by the
H-W group.

Since $T$ is Abelian, the caracterization (\ref{pequegno}) reduces to
(\ref{guenos}), and this leads to the condition:
\bea
(a-1)\frac{\varphi_1}{2\pi} + c\delta\frac{\varphi_2}{2\pi} +
n(-a \frac{x_2}{L_2} +c\delta\frac{x_1}{L_1} - \medio ac\delta) &=& k \in Z\nn\\
(d-1)\frac{\varphi_2}{2\pi} + b\delta^{-1}\frac{\varphi_1}{2\pi} +
n(-b\delta^{-1}\frac{x_2}{L_2} + d \frac{x_1}{L_1} - \medio db\delta^{-1})
&=& k'\in Z \,.
\label{modular}
\eea

\ni With regard to the H-W subgroup (i.e. with $a=d=1$ and $b=c=0$),
we get the same result as in \cite{frachall}: $x_1=k_1\frac{L_1}{n}$
and $x_2=k_2\frac{L_2}{n}$, with $k_1,k_2\in Z$. This implies that
\be
W\equiv \left\{\z(\hat{\eta}_1)^{\frac{k_1}{n}}(\hat{\eta}_2)^{\frac{k_2}{n}},\,\,
k_1,k_2\in Z,\,\, \z\in U(1)\right\} \subset \Gg^{\varphiv} \,,\label{Wilson}
\ee

\ni with $\hat{\eta}_1\equiv e_{(1,0)}$ and $\hat{\eta}_2\equiv e_{(0,1)}$,
for any values of the vacuum angles $\varphi_1$ and $\varphi_2$. These
operators can be interpreted as the Wilson loops in a Chern-Simons theory on
the torus (see section \ref{C-S} and \cite{Poly,Iengo}).

When studying the $SL(2,R)$ subgroup (i.e. with $x_1=x_2=0$), we can proceed
in two ways. Either we can determine for which values of $\varphi_1$ and
$\varphi_2$ we obtain the full modular group $SL(2,Z)$ as good operators, or
we can compute $\Gg^{\varphiv}$ for given values of $\varphi_1$ and $\varphi_2$.

 In the first case, from (\ref{modular}) we easily deduce that modular
invariance is achieved for $\varphi_1=2\pi m_1,
\ \varphi_2=2\pi m_2$ if $n$ is even and for
$\varphi_1=\pi(2m_1+1), \ \varphi_2=\pi(2m_2+1)$ if $n$ is odd, with
$m_1,m_2\in Z$. Clearly, since the vacuum angles are defined modulo $2\pi$
these correspond to $\varphi_1=\varphi_2=0$, or periodic boundary conditions
for $n$ even and to $\varphi_1=\varphi_2=\pi$, or antiperiodic boundary
conditions for $n$ odd.
This is an interesting result, since it reflects the fact that good
operators really depend on the particular representation $D^{\varphiv}$ of
$T$ we are considering.

The group $\Gg$ of good operators for these cases would be obtained by taking the product of
elements of $SL(2,Z)$ with those of $W$ given by (\ref{Wilson}). But from
(\ref{modular}) we see that there are a few more good operators which cannot
be obtained in this way. Altogether, we obtain the
following group of good operators for $\varphi_1=\varphi_2=0$
with $n$ even and for $\varphi_1=\varphi_2=\pi$ with $n$ odd.
:
\be
\Gg = \left\{(S,\frac{1}{n}SJ^m \L_{\k},\z)\ / \ S\in SL(2,Z),\  m=0,1,2,3,\
\k\in Z\times Z,\ \z\in U(1)\right\}
\ee

The computation of $\Gg^{\varphiv}$ for arbitrary values of 
$\varphi_1,\varphi_2$ is a
bit more involved. We have seen that the subgroup $W$ given in (\ref{Wilson})
is always included in $\Gg^{\varphiv}$, so we have only to consider the 
$SL(2,Z)$ subgroup. It is easy to see that if both $\frac{\varphi_1}{2\pi}$ and
$\frac{\varphi_2}{2\pi}$ are irrational, then only the identity matrix in
$SL(2,Z)$ is a good operator, so there is no hint of modular invariance
for this case. If $\frac{\varphi_1}{2\pi}$ is irrational and
$\frac{\varphi_2}{2\pi}=\frac{p}{q}$ is rational (the case obtained by
interchanging 1 and 2 is analogous), then only the subgroup of modular
transformations of the form $\left(\ba{lr} 1 & \epsilon q\delta^{-1}k\\
0&1\ea\right)$ are good operators, with $k\in Z$ and with $\epsilon=1$ for
$n$ even and
$\epsilon=2$ for $n$ odd. If $\frac{\varphi_2}{2\pi}=\frac{p_1}{q_1}$ and
$\frac{\varphi_2}{2\pi}=\frac{p_2}{q_2}$ are rational, then the good operators
are given by the subgroup of modular transformations satisfying the following
diophantine equations:
\bea
(a-1)\frac{p_1}{q_1} + c\delta \frac{p_2}{q_2} - n\frac{ac\delta}{2}
& = & k \in Z \nn \\
b\delta^{-1} \frac{p_1}{q_1} + (d-1)\frac{p_2}{q_2} -n \frac{db\delta^{-1}}{2}
& = & k' \in Z \,.
\eea

\subsection{The Fractional case}
\label{fractional}

For the fractional case, we shall restrict ourselves to the determination
of the subgroup of good operators. The computation of the explicit form
of the constrained wave functions can be performed along the guidelines of the
previous section (they are essentially the ones given in \cite{frachall}
for the H-W group), using the representations of $T$ given in Appendix A.
The dimension of the Hilbert space turns out to be $nr$, and it can be
considered to be a $n$ dimensional Hilbert space made of vector-valued wave
functions, $r$ being the dimension of the vector space.

To determine the subgroup $\Gg$ of good operators, we make use of the
caracterization (\ref{pequegno}) for the litle group, where now, since the
representations are of dimension $r$, the equivalence can be established
through a non-trivial unitary matrix $V(g_g)$.

First, we compute, for
$g_g=(\left(\ba{lr}a&b\\c&d\ea\right),x_1,x_2,\z')\in N_T$,
\bea
g_g*(I_2,k_1L_1,k_2L_2,\z)*g_g^{-1} & =&
(I_2,(ak_1+b\delta^{-1}k_2)L_1,(c\delta k_1+dk_2)L_2,   \nn \\
 & &\z e^{i2\pi\frac{n}{r} [(-a\frac{x_2}{L_2} + c\delta\frac{x_1}{L_1})k_1 +
 (-b\delta^{-1}\frac{x_2}{L_2} + d\frac{x_1}{L_1})k_2]})\,,
\eea

\ni and then we must find for which $g_g\in N_T$ we have
\be
D^{\varphiv}(g_g*g_T*g_g^{-1}) = V(g_g) D^{\varphi}(g_T) V(g_g)^{\dag}
\,\,\,,\,\forall g_T\in T\,,
\ee

\ni where the representations $D^{\varphiv}$ for the fractional case
(obtained in Appendix A), are given by:
\be
D^{\varphiv}(I_2,k_1L_1,k_2L_2,\z) = \z e^{i(\varphi_1k_1+\varphi_2k_2)}
  e^{-i\pi \frac{n}{r}k_1k_2} A_r^{k_1} B_r^{k_2}\,.
\ee

 We proceed as in the integer case, computing firstly the good operators in the
 H-W subgroup. Then the previous equation is written:
\be
e^{i2\pi \frac{n}{r}(-\frac{x_2}{L_2}k_1 + \frac{x_1}{L_1}k_2)}
A_r^{k_1}B_r^{k_2} = V(g_g) A_r^{k_1}B_r^{k_2} V(g_g)^{\dag} \,.
\ee

\ni This equation is the same one which states the equivalence of the
representations $D^{(\mu_1,\mu_2)}$ and $D^{(0,0)}$ and, therefore, making
use of the results given in Appendix A, we find that $x_1=k\frac{L_1}{n}$ and
$x_2=k'\frac{L_2}{n}$, with $k,k'\in Z$. This implies that the subgroup
$W$ given in (\ref{Wilson}) is included in $\Gg^{\varphiv}$ for all values
of $\varphi_1,\varphi_2 \in [0,\frac{2\pi}{r})$.

As far as the $SL(2,Z)$ subgroup is concerned, we shall determine only the
conditions under which full modular invariance is obtained as good operators,
and for this purpose we shall make use of the fact that $SL(2,Z)$ is generated
by two modular transformations:
\be
g_1\equiv \left(\ba{lr}1&1\\0&1\ea\right) \,,\,\,\,\,
g_2\equiv \left(\ba{lr}1&0\\1&1\ea\right)\,.
\ee

\ni Determining under which conditions these two transformations are good
operators will tell us when the theory is fully modular-invariant. For $g_1$ we
obtain the condition:
\be
 e^{i2\pi k_2(\frac{\varphi_1}{2\pi} - \frac{nk_2}{2r})} A_r^{k_1+k_2}
B_r^{k_2} = V(g_1) A_r^{k_1} B_r^{k_2} V(g_1)^{\dag}\,,\,\,\,
  \forall k_1,k_2 \in Z\, .
\ee

\ni For this condition to hold, it is necessary that $\varphi_1=0$ if $nr$ is
even, or $\varphi_1=\frac{\pi}{r}$ if $nr$ is odd. For the first case, the
unitary matrix $V(g_1)$ has the form $V(g_1)_{ij}= \w_r^{\frac{(i-1)^2}{2}}
 \delta_{ij}$, and, for the second, we have
$V(g_1)_{ij}= \w_r^{\frac{i-1}{2n}} \w_r^{\frac{(i-1)^2}{2}} \delta_{ij}$.

For $g_2$ to be a good operator, we obtain the condition:
\be
 e^{i2\pi k_1(\frac{\varphi_2}{2\pi} - \frac{nk_1}{2r})} A_r^{k_1}
B_r^{k_1+k_2} = V(g_2) A_r^{k_1} B_r^{k_2} V(g_2)^{\dag}\,,\,\,\,
  \forall k_1,k_2 \in Z\, .
\ee

\ni Again, for this condition to hold it has to be $\varphi_2=0$ if $nr$ is
even or $\varphi_2=\frac{\pi}{r}$ if $nr$ is odd. The unitary matrix $V(g_2)$
has the form:
\begin{eqnarray}
V(g_2) &=& V = \frac{1}{\sqrt{r}}\left(\ba{ccccc} 1 & \w_r^{\frac{(r-1)(r-2)}{2}}
                             & \ldots &\ldots& 1\\
                         1 & 1 & \ldots & \ldots &  \\
                         \w_r & 1 & \ldots & \ldots & \\
                         \w_r^3 & \w_r & \ldots & \ldots & \vdots\\
                  \vdots & \vdots &\vdots & \vdots & \w_r^{\frac{(r-1)(r-2)}{2}}\\
                   \w_r^{\frac{(r-1)(r-2)}{2}} & \ldots & \ldots &1 & 1
  \ea \right) \hbox{   if $nr$ is even} \nn \\
V(g_2) &=& A_r^{\frac{1}{2n}} V  \hbox{   if $nr$ is odd} \nn \,,
\end{eqnarray}

\ni where $(A_r^{\frac{1}{2n}})_{ij} = e^{i\pi\frac{i-1}{r}}\delta_{ij}$.

It should be stressed that the values of $\varphiv$ for which full modular
invariance is obtained correspond to wave functions which are periodic if
$nr$ is even, or antiperiodic if $nr$ is odd, where these boundary conditions
should be understood with respect to translations by $rL_1$ and $rL_2$.

Note also that the matrix representation $V(g_1)$ and $V(g_2)$ obtained for
$g_1$ and $g_2$ (and therefore for the whole $SL(2,Z)$ group) corresponds to
their action on the $r$-dimensinal vector space. The complete action
of any modular transformation on the wave functions (through $nr\times nr$
matrices) decomposes, thus, in a tensor product of a $n\times n$
matrix and a $r\times r$ matrix, each one acting on different indices
of the wave functions \cite{Iengo}.

This structure of tensor product of the Hilbert space suggests a duality
under the interchange of $n$ and $r$. Indeed, the set of Wilson loops
(\ref{Wilson}) for the theories characterized by $n/r$ and $r/n$ are
isomorphic. Since all the information of the theory is contained in the
Wilson loops, we could say
that the two theories are equivalent. The case $n/r=1$ would, of course,
be self-dual. Moreover, as pointed out in \cite{Imbimbo}, if we denote by
${\cal A}_{\frac{n}{r}}$ the (group) algebra generated by $A$ and $B$
satisfying
\be
AB=e^{i2\pi \frac{n}{r}} BA\,,
\ee

\ni then we have ${\cal A}_{1/(nr)}={\cal A}_{\frac{n}{r}}\times
{\cal A}_{\frac{r}{n}}$. Therefore, the algebra of Wilson loops, besides
being the same for a theory with $T={\cal A}_{\frac{n}{r}}$ and
$T={\cal A}_{\frac{r}{n}}$, is given by the direct product of both
(commuting) algebras. From the point of view of non-commutative $C^*$-algebras,
the algebras ${\cal A}_{\frac{n}{r}}$ and ${\cal A}_{\frac{r}{n}}$ are 
strongly Morita equivalent, which means, in particular, that they possess 
the same representation theory \cite{Rieffel} (see also \cite{Connes}).

\section{$2+1$D Abelian Chern-Simons Theory}
\label{C-S}

As a first application of our results, let us consider a pure topological
field theory on the torus.

Let $M$ be a globally hyperbolic three-dimensional manifold,
 $M=\Sigma\times R$, where  $\Sigma$ is an orientable two-dimentional
manifold.

The action for an Abelian  Chern-Simons (ACS) theory is given by
 \cite{Witten,Poly,Iengo}:
\bea S_{ACS}&=& \frac{k}{4\pi}\int_M \left(A\wedge dA\right)
\label{cs1}\,,
\eea

\ni where   $A$ is a  one-form in $M$ which takes values on the Lie algebra
${\cal K}$ of an Abelian Lie group  $K$. It is straightforward to check
that the action $S_{ACS}$ is invariant under gauge transformations
$A\rightarrow A+ig^{-1}dg$ for any (single-valued) $g:M\longrightarrow K$.

The equations of motion are:
\be d A \equiv F =0 \label{cs4}\,,\ee

\ni the solution of which is the  vector space  ${\cal V}_{ACS}$ of 
all flat connections on $M$. A generic element $A\in{\cal V}_{ACS}$ can be
written in the form $(A_0,ig^{-1}\nabla g + a(t))$, where $a$ is a map from
$R$ to the fibre of $T^*(\Sigma)\otimes {\cal K}$.

This vector space of solutions can be endowed 
with a (pre-)symplectic structure by means of a (pre-)symplectic form
\be
\Omega_{ACS}\left(A',A\right)=\int_\Sigma\>J
= \frac{k}{4\pi}\int_\Sigma\>A'\wedge A\,,
 \label{cs5} \ee

\ni where  $J^\mu\equiv\frac{k}{4\pi}\epsilon^{\mu\nu\sigma}{A'}_\nu A_\sigma$
is a divergenceless current which ensures the independence of
$\Omega_{ACS}\left(A',A\right)$ on the chosen Cauchy factorization of $M$, 
$M=\Sigma\times R$.
 
Since the exterior derivative $d$ commutes with the pullback operator $*$,
if  $f$ is a diffeomorphism of  $M$,  and
$A',A\in {\cal V}_{ACS}$, then  $A'+f^*A$ is also a solution of 
(\ref{cs4}).

With this information, we can propose a quantizing group 
$\widetilde{G}_{ACS}$
for this theory, the composition law of which is:
\bea f''&=&f'\circ f\>,\quad f,f',f''\in {\rm Diff}({M})\nonumber\\
A''&=&{f^{-1}}^*A'+A\label{tGACS}\\
\zeta''&=&\zeta\zeta'\exp\Omega_{ACS}\left({f^{-1}}^*A',A\right)\,,
\nonumber\eea

\ni i.e.,  the extension by $U(1)$ of the semidirect product
${\cal V}_{ACS}\otimes_s{\rm Diff}(M)$. The characteristic subgroup 
(generated by the kernel of $\Omega_{ACS}$, see Sec. \ref{Schrodinger}) of 
this group proves to be  $G_{\Omega}= \{(f,A,1)/\,
A=(A_0,ig^{-1}\nabla g)\quad\hbox{for some}\quad g: M\rightarrow
K\}\subset\widetilde{G}_{ACS}$, which contains the {\it gauge group} 
$G_{\rm gauge}$ of the theory, constituted by all (single-valued) 
$g:M\rightarrow K$ [To be precise, $G_{\rm gauge}$ has not been included in 
$\widetilde{G}_{ACS}$, but rather an orbit of it under the action 
$A\rightarrow A+ig^{-1}dg$. Including the group  $G_{\rm gauge}$ explicitly 
in $\widetilde{G}_{ACS}$ requires a slight modification of the notion of 
gauge transformation \cite{Empro}. For the present case, we shall assume the 
existence of a group $G_{\rm gauge}$ and a subgroup of $\widetilde{G}_{ACS}$ 
directly related to it]. Thus,
the polarization conditions (which contains the characteristic subgroup)
imply that wave functions depend only on topological
and gauge invariant quantities. For this kind of theory, standard approaches 
claim that all gauge-invariant information of a connection can be extracted
from the {\it Wilson loops} defined by:
\be W(A,\gamma)=\exp \int_\gamma A\label{wilsonloops}\,,
\ee

\ni for any loop   $\gamma$ in $\Sigma$. Since connections $A$ are flat, the
 Wilson loops will depend only on the homotopy class
 $[\gamma]\in \pi_1(\Sigma)$ of the corresponding loop $\gamma$. For this 
reason the normal subgroup $Diff_0(M)\subset Diff(M)$ of diffeomorphism 
of $M$ connected to the identity acts trivially on the Wilson loops. Therefore
the diffeomorphisms that really matter in (\ref{tGACS}) are the quotient
$Diff(M)/Diff_0(M)$ called the {\it modular group} 
(see \cite{Martin,Nakahara}) of the Riemann surface
$\Sigma$ (note that all diffeomorphisms of the $R$ part of $M$ are connected 
to the identity).     

It should be stressed that if $\pi_1(\Sigma)=0$ (what implies that 
$H^1(\Sigma)=0$) then the ACS theory is trivial since all connections are of 
the form $A=ig^{-1}dg$ for some 
(always single-valued) $g:M \rightarrow K$. This implies that 
$G_{\Omega}=G_{ACS}\equiv {\cal V}_{ACS}\otimes_s{\rm Diff}(M)$ and
$\widetilde{G}_{ACS} = G_{ACS}\times U(1)$, that is, the central extension is
trivial (another way of seeing this is that since the homotopy group of 
$M$ is trivial, all Wilson loops are trivial). Therefore we assume that 
$\Sigma$ is a multiply-connected oriented
two-dimensional manifold. What makes the theory non-trivial in this case is 
the fact that the gauge group, $G_{\rm gauge}$, is smaller than its 
simply-connected counterpart in the universal covering space of $M$, 
${\bar G}_{\rm gauge}$, constituted by all ${\bar g}:{\bar M}\rightarrow K$, 
with ${\bar M}={\bar \Sigma}\times R$, and ${\bar \Sigma}$ the universal 
covering space of $\Sigma$. In fact, the group
$G_{\rm gauge}\subset{\bar G}_{\rm gauge}$ is made of those elements 
${\bar g}\in{\bar G}_{\rm gauge}$ verifying 
${\bar g}\circ [\gamma] = {\bar g}$, where here $[\gamma]$ represents the 
natural action (as diffeomorphism) of 
the homotopy class $[\gamma]\in \pi_1(M)$ on ${\bar M}$.

For the present case, $\Sigma=S^1\times S^1$ and $K=U(1)$. The space 
${\cal V}_{ACS}$ is made of connections of the form 
$(A_0,ig^{-1}\nabla g+a(t))$, where $g$ is single-valued on the torus.
The solution manifold, that which remains once the quotient by the
characteristic subgroup $G_{\Omega}$ is taken, is parameterized by the variables
$a_1(t),a_2(t)$ modulo an integer, defining a torus. 
The reason is that $G_{\Omega}$ also contains the {\it global (large) gauge 
transformations} (see for instance \cite{Iengo} and references therein),
\be
a_j\rightarrow a_j+k_j\,,\;k_j \in Z\label{gglobal}\,,
\ee

This large gauge transformations are clearly seen to come from transformations
of the form $g=exp(ik_jx^j)$, with $k_j\in Z$ and $\{x^j\}$ a set of local 
coordinates on the torus, in such a way that $0$ and $2\pi$ are identified. 
The reason of the restriction of $k_j$ to integers is the condition  
${\bar g}\circ [\gamma] = {\bar g}$. This indicates that the gauge group 
$G_{\rm gauge}$ is a disconnected group, with 
$G_{\rm gauge}/G_{{\rm gauge}}^0= Z\times Z$, $G_{{\rm gauge}}^0$ being the 
connected component of the identity.
  
For this reason, in the quantum theory the operator associated with the 
variables $A$ (more precisely, $a$) are not properly defined (they are 
{\it bad operators}, see Sec. \ref{AQG}). We must resort to single-valued 
({\it good}) operators of the form:
\be
W(A)=e^{2\pi i \sum_{j=1}^{2} n_j a_j}\label{wily}\,,
\ee

\ni where $n_j\in Z$ should be interpreted as the winding number of a path 
$\gamma$ around the cycle $j$. Remember that for the torus, the homotopy 
classes $[\gamma]$ are generated by two elements, $[\gamma_j]\;j=1,2$, 
representing loops (with winding number one) around each one of the two 
cycles of the torus. The modular group proves to be 
$Diff(T^2)/Diff_0(T^2)=SL(2,Z)$. 

At this point it should be stressed that the resulting theory corresponds to 
a quantum mechanical system with phase space a torus parameterized by 
$(q,p)\equiv (a_1 \hbox{ mod } 1,a_2 \hbox{ mod } 1)$.

According to this equivalence, we could have studied this system in the 
framework of AQG by starting with the group H-W$\otimes_s Diff(R^2\times R)$
with structure group $T$ a fibre bundle with base $Z\times Z$ and fibre 
$U(1)$, where $Z\times Z$ is the subgroup of $Diff(R^2\times R)$ of 
translations by $(k_1L_1,k_2L_2)$, with $k_1,k_2\in Z$. Since the 
only relevant diffeomorphisms at the final theory on the torus will be the 
modular transformations $SL(2,Z)\subset SL(2,R)$, it is enough to start with
H-W$\otimes_s SL(2,R)$, which is the Schr\"odinger group. Thus, all the results
of Sec. \ref{Schrodinger-toro} apply here. Of course, we could have 
started directly with H-W$\otimes_s SL(2,Z)$, but this group, being 
disconneted, is more difficult to quantize than the Schr\"odinger group 
(in particular, finding a polarization for this group is a difficult task). 
In addition, we think that showing how $SL(2,Z)$ emerges as good operators 
is a very illustrative way of studying the problem.


\vskip 0.5cm
{\it In summary}:

\begin{itemize}
\item The coupling constant  $k$ plays the same role as the quantity
 $\phi\equiv\frac{m\w L_1L_2}{2\pi\hbar}$ in the Schr\"{o}dinger group on the
 torus, determining the character of the resulting (finite-dimensional)
 Hilbert space.
\item The set of Wilson loops (\ref{wily}) takes part of the set of
good operators in our language. More precisely, they are the analogue of
the set $W$ given in (\ref{Wilson}).
\item The group of large gauge transformations is the analogue of the
structure group $T$. When the coupling constant $k$ is fractional, this gauge
group is called {\it anomalous} \cite{Iengo} because of its non-Abelian
character due to the non-trivial fibration for this case, as oposed to the
original Abelian gauge group $K$.
\item The non-equivalent representations of $T$, parametrized by the indices
$\varphi_1,\varphi_2$ (vacuum angles), characterize the non-equivalent
quantizations of the theory.
\end{itemize}

The Chern-Simons theory constitutes a particular example of a drastic
reduction of the number of  original infinite (field) degrees of freedom to a
finite number (which, in addition, contain a finite number of states, due to
the compactness of the phase space when restricted to the torus), as a
consequence of a huge gauge invariance which kills all of them except for
the topologic ones.

\subsection{Further comments}

Comparing our results with those in the literature, we find full agreement
with  Ref. \cite{Poly}, in the context of $U(1)$ Chern-Simon theory on the 
torus, as far as the integer case is concerned. For the fractional case,  
an apparent discrepance  with the results in \cite{Poly} appears: in our 
notation, modular invariance is obtained only for $n$ even (and any value of 
$r$) and vacuum 
angles $\varphi_1=\varphi_2=0$. However, the agreement is achieved if we 
realize
that the proper range of inequivalent vacuum angles in \cite{Poly} should be
$[0,\frac{2\pi}{r})$. 


This problem was also studied in Ref. \cite{Iengo} (also in the context of
$U(1)$ Chern-Simons theory and anyons on the torus), where full modular
invariance was obtained for both the integer and fractional case, but they
claimed that the vacuum angles always
have to be  $\medio$ disregarding the parity of the coupling constant
(the equivalent of our $\frac{n}{r}$). A more detailed
analysis of their results reveals that the vacuum angles they introduce are
defined modulo $\frac{1}{nr}$, and $(\medio\  {\rm mod}\
\frac{1}{nr})$ is $0$
for even $nr$ and $\frac{1}{2nr}$ for odd $nr$, corresponding to periodic and
antiperiodic boundary conditions, respectively. Therefore, their results
completely agree  with ours.

In \cite{Imbimbo}, a non-Abelian Chern-Simons theory is considered, with
gauge group $SL(2,R)$. When restricted to the torus, they obtained
essentially the same results as ours and those of \cite{Poly,Iengo} (with 
the abovementioned remarks) with
respect to the Hilbert space and the set of observables (good operators),
because the reduced phase space of the theory is almost the space of
flat connections of an Abelian gauge group.

\section*{Aknowlodgements}

J. Guerrero thanks the Spanish MEC for a Postdoctoral grant and the Department
of Physics of Naples-INFN for its hospitality and financial support, 
and M. Calixto thanks the Spanish MEC for a FPI grant. \Comision

\appendix

\section{Appendix: Unitary and Irreducible representations of $T$}

The structure subgroup $T$, as defined in section \ref{Schrodinger-toro}, is 
a $U(1)$ bundle with base $\Gamma_{\L}$, and can be written as:
\be
T=\{(I_2,k_1L_1,k_2L_2,\z)\in \Gtm\,/\, k_1,k_2\in Z\,,\,\z\in U(1)\}\,,
\ee

\ni with group law derived from the group law of the Schr\"{o}dinger group:
\be
(I_2,k_1'L_1,k_2'L_2,\z')*(I_2,k_1L_1,k_2L_2,\z)=
  (I_2,(k_1'+k_1)L_1,(k_2'+k_2)L_2,
   \z'\z e^{i\frac{m\w L_1L_2}{2\hbar}(k_1'k_2-k_2'k_1)})\,.
\label{leyT}
\ee

To determine the structure of $T$, we compute the group commutator of two
elements:
\be
[(I_2,k_1'L_1,k_2'L_2,\z'),(I_2,k_1L_1,k_2L_2,\z)] =
  (I_2,0,0,e^{i\frac{m\w L_1L_2}{\hbar}(k_1'k_2-k_2'k_1)})\,,
\ee

\ni from which we see that its structure depends on the value of
$\frac{m\w L_1L_2}{2\pi\hbar}$, in such a way that there are three
possibilities:
\begin{itemize}
\item[$i)$] {\sc integer case}: $\,\,\frac{m\w L_1L_2}{2\pi\hbar} = n \in Z$.
\item[$ii)$] {\sc fractional case}: $\,\,\frac{m\w L_1L_2}{2\pi\hbar}=
\frac{n}{r}\,,\,\,n,\,r \in Z$ and relative prime (with $r > 1$).
\item[$iii)$] {\sc irrational case}: $\,\,\frac{m\w L_1L_2}{2\pi\hbar}= \rho$,
with $\rho$ an irrational number.
\end{itemize}

Let us study the integer and fractional case separately. The irrational case 
will not be considered here (see \cite{Connes} for a detailed study of this 
case).

\vskip 0.5cm
\ni {\sl INTEGER CASE:}
\vskip 0.3cm

 In this case, $T$ is an Abelian group, and therefore
 $T=\Gamma_{\L} \times U(1)$ and all its representations are of
 dimension 1. As stated above, we shall consider only those
 representations, which restricted to $U(1)$, are the natural representations,
 and these have the form:
 \be
 D^{\varphiv}(I_2,k_1L_1,k_2L_2,\z)=\z e^{i(\varphi_1k_1+\varphi_2k_2)}
 e^{-i\pi nk_1k_2}\,, \,\,\,\,\,\forall k_1,k_2\in Z,\,\,\forall \z\in U(1),\,
 \ee

\ni where the range of inequivalent representations, since they are
1-dimensional, is given simply by $\varphi_1,\varphi_2 \in [0,2\pi)$. Note
that, except for the term $e^{-i\pi nk_1k_2}$, this is the product of the
natural representation of $U(1)$ times a representation of
$\Gamma_{\L}\approx Z\times Z$. This extra term is only a
coboundary coming from the fact that we have used Bargmann's cocycle in the
group law
of the Schr\"{o}dinger group, and Bargmann's cocycle does not satisfy the
conditions given in \cite{frachall} for the possible cocycles for the
H-W group on the torus. Note, thus, that this restriction can be relaxed
by introducing this coboundary term in the representations of $T$.

\vskip 0.5cm
\ni {\sl FRACTIONAL CASE:}
\vskip 0.3cm

In this case, $T$ is not Abelian, and the commutator of two elements has the
form:
\be
[(I_2,k_1'L_1,k_2'L_2,\z'),(I_2,k_1L_1,k_2L_2,\z)] =
  (I_2,0,0,\w_r^{k_1'k_2-k_2'k_1})\,,
\ee

\ni where $\w_r\equiv e^{i2\pi\frac{n}{r}}$ is a $r^{th}$ root of unity. Note
that if $|n|>r$, then $w_r=e^{i2\pi\frac{n}{r}}=e^{i2\pi\frac{q}{r}}$,
where $q=n \ {\rm mod} \ r$. Since $n$ and $r$ are relative prime,
$q$ and $r$ turn out also to be relative prime and, therefore, we can use
either of the two pairs to characterize $T$.

 The group $T$ admits a non-trivial characteristic subgroup
 (see \cite{frachall}), of the form:
\be
G_C = \{(I_2,rk_1L_1,rk_2L_2,e^{i\pi nrk_1k_2})\,/\, k_1,k_2\in Z\}\,.
\ee

\ni The characteristic subgroup can be identified in this case with the Casimir
elements of $T$ which are not in $U(1)$, i.e. those elements of $T$ (not 
belonging to $U(1)$) which commute with all other elements in $T$. In fact, 
the centre of $T$ is given by $G_C\times U(1)$.

If we quotient $T$ by $G_C$, we obtain a group which is a {\it generalized
Clifford group} $G_2^r$ (see \cite{Clifford} for the definition and the study
of representations of generalized Clifford groups) times $U(1)$.
Therefore, the representations of $T$ can be obtained from those of $G_C$
and $G_2^r$ (and the natural representation of $U(1)$).

The representations of $G_C$, being isomorphic to $Z\times Z$, are
characterized by two ``vacuum angles'' $\varphi_1,\varphi_2$, whose range
of non-equivalence should be determined. The representations of $G_2^r$
are studied in detail in \cite{Clifford}, so that here we shall give
only the results. It should be remarked, however, that in \cite{Clifford}
$\w_r$ is an arbitrary $r^{th}$ root of unity, and different choices for
it give inequivalent representations of $G_2^r$, whereas here the value
of $\w_r$ is given {\it a priori} (it is determined by the fact that
$T$ is a subgroup of \Gt), so that the representation of $G_2^r$ is
uniquely determined. In addition, since $n$ and $r$ are relative prime,
$\w_r$ is a primitive $r^{th}$ root of unity, implying that the
representation of $G_2^r$ associated with it is of dimension $r$, either
for prime or non-prime $r$ .

 The $r$-dimensional unitary irreducible representation of $G_2^r$ can be
constructed with the aid of two $r\times r$ matrices, $A_r$ and $B_r$:
\be
(A_r)_{ij} = \w_r^{i-1} \delta_{ij} \,\,,\,\,
(B_r)_{ij}=\delta_{i,(j\ {\rm mod}\ r)+1} \,,\,\,\,\, i,j=1,2,...,r\,,
\ee
\ni verifying  $A_rB_r = \w_r B_rA_r$, and $A_r^r=B_r^r=I_r$. Putting together
this representation of $G_2^r$ and that  of $G_C\approx Z\times Z$,
we can build a representation for the entire $T$, of the form:
\be
D^{\varphiv}(I_2,k_1L_1,k_2L_2,\z) = \z e^{i(\varphi_1k_1+\varphi_2k_2)}
  e^{-i\pi \frac{n}{r}k_1k_2} A_r^{k_1} B_r^{k_2}\,,\,\,\,\,\,
\forall k_1,k_2\in Z,\,\,\forall \z\in U(1)\,. \label{repTfrac}
\ee

One would na\"{\i}vely expect that the range of non-equivalent representations
would be $\varphi_1,\varphi_2\in [0,2\pi)$, as in the integer case. However,
since the representations are not 1-dimensional, there could be non-trivial
unitary transformations relating representations in this interval and,
therefore, reducing the range on non-equivalence.

Thus, we have to determine the minimum values of $\mu_1,\mu_2$ for which the
representation $D^{(\mu_1,\mu_2)}$ is equivalent to the trivial
representation $D^{(0,0)}$; i.e. there exists a unitary matrix $V$ such that
$D^{(\mu_1,\mu_2)} = V D^{(0,0)} V^{\dag}$. Studying separately the cases
$(\mu_1,0)$ and $(0,\mu_2)$, and after a few computations, we obtain:
\begin{itemize}
\item[(i)] $D^{(\mu_1,0)}$ is equivalent to $D^{(0,0)}$ for
             $\mu_1=\frac{2\pi}{r}$, with $V=B_r^{m_0}$.
\item[(ii)] $D^{(0,\mu_2)}$ is equivalent to $D^{(0,0)}$ for
             $\mu_2=\frac{2\pi}{r}$, with $V=A_r^{\frac{1}{n}}$.
\end{itemize}

Here $(A_r^{\frac{1}{n}})_{ij}\equiv \w_r^{\frac{i-1}{n}}\delta_{ij} =
e^{i2\pi \frac{i-1}{r}} \delta_{ij}$, and $0<m_0<r$ is an integer solution
of the diophantine equation:
\be
1+n m_0 = r k \,\,, k\in Z \,,
\ee

\ni which has always two solutions in the range
$\{-(r-1),\ldots,0,\ldots,r-1\}$, provided $n$ and $r$ are relative prime
(this is a particular case of the Bezout lemma, for $gcd(n,r)=1$, which
in turn can be proven using Euclidean division of integers, see, for instance,
\cite{Algebra}).
Note that $(A_r^{\frac{1}{n}})^n = A_r$ and $(B^{m_0})^n =B_r^{-1}$, so
that these matrices can be considered as the $n^{th}$ roots of the
matrices $A_r$ and $B_r^{-1}$, respectively.

Therefore, the range of non-equivalent representations of $T$ is reduced
to $\varphi_1,\varphi_2 \in [0,\frac{2\pi}{r})$. This fact will be of
extreme importance for the determination of the good operators in the
fractional case.

\end{document}